\documentclass[a4paper,10pt,aps,amsfonts,amsmath,nofootinbib,byrevtex,prd,notitlepage]{revtex4-2}

\usepackage[UKenglish]{babel}
\usepackage{drcmath}
\usepackage{graphicx}

\usepackage[utf8]{inputenc}

\DeclareVector{\valpha}{\alpha}

\begin{document}

\title{The Equal-Time Quark Propagator in Coulomb Gauge}

\author{Davide Campagnari}
\author{Hugo Reinhardt}
\affiliation{Institut f\"ur Theoretische Physik, Universit\"at T\"ubingen,
Auf der Morgenstelle 14, 72076 T\"ubingen, Germany}
\date{\today}


\begin{abstract}
We investigate the equal-time (static) quark propagator in Coulomb gauge within
the Hamiltonian approach to QCD. We use a non-Gaussian vacuum wave functional which
includes the coupling of the quarks to the spatial gluons. The expectation value
of the QCD Hamiltonian is expressed by the variational kernels of the vacuum wave
functional by using the canonical recursive Dyson--Schwinger equations (CRDSEs)
derived previously. Assuming the Gribov formula for the gluon energy we solve the CRDSE
for the quark propagator in the bare-vertex approximation together with the
variational equations of the quark sector. Within our
approximation the quark propagator is fairly insensitive to the
coupling to the spatial gluons and its infrared behaviour is exclusively determined by the
strongly infrared diverging instantaneous colour Coulomb potential.
\end{abstract}

\maketitle


\section{Introduction}

Confinement and the spontaneous breaking of chiral symmetry leading to the dynamical
generation of a constituent quark mass are the most important low-energy
phenomena of Quantum Chromodynamics (QCD) at ordinary density and temperature.
Research on this subject has been ongoing for decades
(see Refs.~\cite{Alkofer:2000wg,Pawlowski:2005xe,Fischer:2006ub,Binosi:2009qm,Braun:2011pp,Eichmann:2016yit,Aoki:2016frl}
for some recent reviews) and various pictures have emerged, like the dual Mei\ss ner
effect \cite{Nambu:1974zg,tHooft:1981bkw}, the center vortex scenario
\cite{Mack:1978rq,Nielsen:1979xu,DelDebbio:1998luz,Langfeld:1997jx,Engelhardt:1999fd}, and the
Kugo--Ojima--Gribov--Zwanziger scenario \cite{Gribov:1977wm,Kugo:1979gm,Zwanziger:1998ez,Feuchter:2004mk,Epple:2006hv}.
Furthermore, these pictures are all supported by lattice calculations \cite{Greensite:2011zz}
and have been shown to be closely related \cite{Greensite:2004ur,Reinhardt:2008ek,Burgio:2015hsa}.
Despite these efforts, a rigorous understanding of these phenomena is still lacking.

Much progress has been made in recent years using functional continuum
methods like Dyson--Schwinger equations (DSEs)
\cite{Alkofer:2000wg,Fischer:2006ub,Binosi:2009qm},
Functional Renormalization Group (FRG) flow equations
\cite{Pawlowski:2005xe,Braun:2011pp},
or variational methods in the Hamiltonian
\cite{Schutte:1985sd,Szczepaniak:2001rg,Feuchter:2004mk,Epple:2006hv,Campagnari:2010wc,%
Pak:2011wu,Pak:2013uba,Campagnari:2015zsa,Vastag:2015qjd,Reinhardt:2016pfe,Campagnari:2016wlt,Campagnari:2018flz}
or Lagrange \cite{Quandt:2013wna,Quandt:2015aaa} form.
The variational approach has, in principle, the advantage over other continuum
methods that it provides a criterium for the improvements of (the variational
ansatz and) the truncations used. However, when one goes beyond Gaussian wave
functionals 
one faces a problem: Wick's
theorem no longer applies, making a direct evaluation of expectation values in
terms of the variational kernels of the vacuum wave functional impossible. In
Refs~\cite{Campagnari:2010wc,Campagnari:2015zsa} this problem was elegantly solved
using DSE techniques. The upshot is a set of DSE-like equations (named canonical
recursive Dyson--Schwinger equations, CRDSEs) which relate the various $n$-point functions of the quark and
gluon fields to the variational kernels of the vacuum wave functional.
In the present paper we investigate the CRDSE for the (static) quark propagator.

The organization of the paper is as follows: In Sec.~\ref{sec-def} we give a short
summary of the Hamiltonian approach to QCD in Coulomb gauge
developed in Ref.~\cite{Campagnari:2015zsa}. In Sec.~\ref{sec-trunc}
we discuss the truncation and renormalization scheme of the quark propagator CRDSE.
The numerical input and the results are presented in Sec.~\ref{sec-res}, and some
final remarks are given in Sec.~\ref{sec-conc}. In the Appendix we present the
details of the infrared analysis of the relevant CRDSEs.


\section{Coulomb Gauge Equal-Time Quark Propagator}
\label{sec-def}

In terms of the fermion field operators $\psi$, $\psi^\dag$ the equal-time
quark propagator $S$ is defined by
\begin{equation}\label{qp1}
S(1,2) \coloneq \frac12 \vev*{\comm{\psi(1)}{\psi^\dag(2)}} .
\end{equation}
We use here a condensed notation where a single numerical index stands collectively
for the spatial coordinate and the colour and spinor indices. A repeated numerical
label [see e.g.~\Eqref{coulkernel} later] implies integration over the spatial coordinate as well as summation over all
discrete indices.
Both the factor $1/2$ and the commutator on the right-hand side of \Eqref{qp1}
arise from the equal-time limit of the time-dependent propagator; the expectation
value $\vev{\ldots}$ is taken in the ground state of the QCD Hamiltonian, which
in Coulomb gauge takes the form \cite{Christ:1980ku}
\begin{equation}\label{hh1}
\begin{split}
H ={}& \frac12 \int\d^3x \, J_A^{-1} E_i^a(\vx) J_A \, E_i^a(\vx) + \frac12 \int\d^3x \, B_i^a(\vx) \, B_i^a(\vx) \\
    &+ \int\d^3x \, \psi^\dag(\vx) \bigl[ -\I*\valpha\cdot\grad - g \valpha\cdot\vec{A}(\vx) + \beta m \bigr] \psi(\vx) \\
    &+ \frac{g^2}{2} \int \d^3x \d^3y \, J_A^{-1} \rho^a(\vx) \,J_A \, F_A^{ab}(\vx,\vy) \, \rho^b(\vy) .
\end{split}
\end{equation}
Here, $E_i^a=\I*\delta/\delta A_i^a$ and $B_i^a$ are the chromoelectric and -magnetic fields, $\alpha_i$
and $\beta$ are the usual Dirac matrices, $m$ is the bare current quark mass,
and $\vec{A}=\vec{A}^{\!a\,} t^a$ are the (transverse) gauge fields with
$t^a$ being the hermitian generators of the $\mathfrak{su}(\Nc)$ algebra.
The last term in \Eqref{hh1} is the so-called Coulomb term: it describes the
interaction of the colour charge density
\[
\rho^a = \psi^\dag \, t^a \psi - f^{abc} A_i^b \, E_i^c
\]
through the Coulomb kernel
\begin{equation}\label{coulkernel}
F_A(1,2) = G_A(1,3) \, G_0^{-1}(3,4) G_A(4,2) ,
\end{equation}
where
\[
G_A^{-1}(\vx,\vy) = \bigl( - \delta^{ab} \nabla^2_x - g f^{acb} A_i^c(\vx) \partial_i^x \bigr) \delta(\vx-\vy), \qquad G_0 \equiv G_{A=0}
\]
is the Faddeev--Popov operator of Coulomb gauge. Finally, $J_A=\Det G_A^{-1}$ 
is the corresponding Faddeev--Popov determinant.

For the fermionic operators $\psi$, $\psi^\dag$ we use a representation based on coherent states $\ket{\xi}$ 
(see Ref.~\cite{Campagnari:2015zsa} for details), where the action
of $\psi$, $\psi^\dag$ onto a fermionic state $\ket{\varPhi}$ reads
\begin{equation}\label{csr1}
\bra{\xi} \psi(1) \ket{\varPhi} = \biggl( \xi_-(1) + \frac{\delta}{\delta \xi_+^\dag(1)} \biggr) \varPhi[\xi] , \qquad
\bra{\xi} \psi^\dag(1) \ket{\varPhi} = \biggl( \xi_+^\dag(1) + \frac{\delta}{\delta \xi_-(1)} \biggr) \varPhi[\xi] .
\end{equation}
In \Eqref{csr1}, $\varPhi[\xi] \equiv \braket{\xi}{\varPhi}$ is the coherent-state
functional representation of the state $\ket\varPhi$, and
\[
\xi(1) = \xi_+(1) + \xi_-(1) , \quad \xi_\pm(1) = \Lambda_\pm(1,2) \xi(2)
\]
is a spinor-valued Grassmann field, with
\begin{equation}\label{qptproj3}
\Lambda_\pm (1, 2) = \int \dfr[3]{p} \e^{\I* \vp \cdot (\vx_1 - \vx_2)} \Lambda_\pm (\vp) ,
\qquad \Lambda_\pm(\vp) = \frac12 \pm \frac{\valpha\cdot\vp+\beta m}{2 \sqrt{\vp^2+m^2}}
\end{equation}
being the projectors onto positive/negative energy eigenstates of the free Dirac
operator.

Denoting the ground state of the QCD Hamiltonian by $\varPsi[\xi,A]$, the vacuum
expectation value $\vev{O}$ of an operator $O$ depending on both the gauge field~$A$
and the fermionic fields $\psi^\dag$, $\psi$ is given by the functional integral
\begin{equation}\label{evg1}
\vev{O\bigl[A,E,\psi,\psi^\dag\bigr]}
= \int \calD \xi^\dag \, \calD \xi \, \calD A \, J_A \e^{-\mu} \, \varPsi^*[\xi,A] \,
 O \biggl[A,\I \frac{\delta}{\delta A}, \xi_-+\frac{\delta}{\delta\xi_+^\dag}, \xi_+^\dag+\frac{\delta}{\delta\xi_-} \biggr] \,
\varPsi[\xi,A] ,
\end{equation}
where
\[
\mu = \xi^\dag(1) \, S_0(1,2) \, \xi(2)
\]
is the integration measure of the coherent fermion states, which involves the bare
quark propagator
\begin{equation}\label{evg4}
S_0(1,2) = \frac{1}{2} \bigl[ \Lambda_+(1,2) - \Lambda_-(1,2) \bigr] ,\qquad
S_0(\vp) = \frac{\valpha\cdot\vp + \beta m}{2 \sqrt{\vp^2+m^2}} .
\end{equation}
The vacuum wave functional $\varPsi[\xi,A]$ can be written as
\begin{equation}\label{vwf}
\varPsi[\xi,A] \propto \exp\biggl\{ -\frac12 \, S_A[A] - S_f [\xi,A] \biggr\},
\end{equation}
where $S_A$ defines the vacuum wave functional of the pure Yang--Mills theory,
while $S_f$ is the fermionic contribution, which includes also the coupling of
the quarks to the spatial gluons. When all functional derivatives in \Eqref{evg1}
are worked out, expectation values of operators reduce to quantum averages of
field functionals with an ``action''
\begin{equation}\label{hh5}
S = S_A + S_f^{} + S_f^* + \mu .
\end{equation}
This formal equivalence between vacuum expectation values in the Hamiltonian
approach and functional integrals of a Euclidean field theory can be exploited
to derive exact functional equations similar to Dyson--Schwinger equations \cite{Campagnari:2010wc,Campagnari:2015zsa}.
They differ from the standard DSEs in the sense that they do not connect
propagators and vertices with the ordinary action but rather
with the exponent of the vacuum wave functional.
To stress the conceptual difference between the usual DSEs and the equations
arising in our approach we named these `canonical recursive DSEs' (CRDSEs).

The vacuum wave functional \Eqref{vwf}, and hence the action \Eqref{hh5},
is eventually determined by means of the variational principle: first we choose
a suitable Ansatz for the vacuum wave functional, which depends on some variational
kernels whose high-energy behaviour is known from perturbation theory
\cite{Mansfield:1997kj,Mansfield:1998vc,Campagnari:2009km,Campagnari:2009wj,Campagnari:2014hda},
unless they are of non-perturbative origin.
Then we evaluate the vacuum expectation value $\vev{H}$ of the Hamiltonian using
the CRDSEs to relate the $n$-point functions appearing in $\vev{H}$ to the variational
kernels of the vacuum wave functional. The resulting vacuum energy is then
minimized with respect to the variational kernels. This procedure results in a
set of gacoupledp equations for the variational kernels, which have to be solved
together with the CRDSEs. The Ansatz chosen in
Refs.~\cite{Vastag:2015qjd,Campagnari:2016wlt,Campagnari:2018flz} reads
\[
S_f + S_f^* = \xi^\dag(1) \, \bar\gamma(1,2) \, \xi(2) + \xi^\dag(1) \, \bar\Gamma_0(1,2;3) \, \xi(2)A(3) ,
\]
where the biquark kernel
\begin{equation}\label{ans10a}
\bar\gamma(1,2) = \Lambda_+(1,1') \, K_0(1',2') \, \Lambda_-(2',2) +  \Lambda_-(1,1') \, K_0^\dag(1',2') \, \Lambda_+(2',2)
\end{equation}
and the bare quark-gluon vertex
\begin{equation}\label{qgv0}
\bar\Gamma_0(1,2;3) = \Lambda_+(1,1') \, K(1',2';3) \, \Lambda_-(2',2) +  \Lambda_-(1,1') \, K^\dag(1',2';3) \, \Lambda_+(2',2)
\end{equation}
involve the variational kernels $K_0$ and $K$, for which we have chosen the form
\begin{align}
\label{sk1}
K_0(\vp) &= \beta s(\vp) , \\
\label{vk1}
K_i^{a}(\vp,\vq;\vk) &= g t^a \bigl[ \alpha_i \, V(\vp,\vq) + \beta \alpha_i \, W(\vp,\vq) \bigr] (2\pi)^3 \delta(\vp+\vq+\vk) .
\end{align}
A more general form for $K_0$ could be chosen, but as shown in Ref.~\cite{Campagnari:2018flz}
the simple form \Eqref{sk1} already captures the relevant physics. The variational
kernel $K$ [\Eqref{vk1}] contains a leading-order term involving the Dirac matrix
$\alpha_i$ and a further contribution proportional to $\beta \alpha_i$. The latter
term, introduced in Ref.~\cite{Vastag:2015qjd}, turns out to be of purely
non-perturbative nature but is crucial to ensure one-loop renormalizability
of the physical quark propagator \cite{Campagnari:2018flz}.
In Refs.~\cite{Vastag:2015qjd,Campagnari:2016wlt,Campagnari:2018flz} the
vacuum expectation value of the QCD Hamiltonian was calculated in the chiral
limit $m=0$ up to including two-loop order,
and the minimization with respect to the variational kernels resulted in four
coupled equations for the scalar functions $s$, $V$, $W$, and the gluon energy $\Omega$.
The equations for the vector kernels $V$ and $W$ can be explicitly solved in terms
of the scalar kernel $s$ and the gluon energy $\Omega$, yielding\footnote{%
    Whenever there is no ambiguity,  we will write the momentum dependence as a
    subscript in order to simplify the notation.}
\begin{subequations}\label{vk}
\begin{align}
\label{vkV}
V(\vp,\vq) &= - \frac{1 + s_p s_q}{\Omega(\vp+\vq) + \abs{\vp} \frac{1-s_p^2+2s_p s_q}{1+s_p^2} + \abs{\vq} \frac{1-s_q^2+2s_p s_q}{1+s_q^2}} ,\\
\label{vkW}
V(\vp,\vq) &= - \frac{ s_p - s_q }{\Omega(\vp+\vq) + \abs{\vp} \frac{1-s_p^2-2s_p s_q}{1+s_p^2} + \abs{\vq} \frac{1-s_q^2-2s_p s_q}{1+s_q^2}} .
\end{align}
\end{subequations}
The scalar kernel $s_p$ itself obeys a gap equation (see Sec.~\ref{sec:sk} later).
The same is true for the gluon energy $\Omega$. It was shown in Ref.~\cite{Vastag:2015qjd}
that the unquenching of the gluon energy by the quarks is negligible. We will therefore
use for $\Omega$ the Gribov formula, see \Eqref{gribov} below, which nicely fits
the lattice data for the gluon propagator in pure Yang--Mills theory.

As pointed out above, in the CRDSEs the variational kernels
occurring in the vacuum wave functional [or, more exactly, their combinations
Eqs.~\eqref{ans10a} and \eqref{qgv0}] play the role of the bare vertices
in the Lagrangian approach. We
stress, however, that the word `bare' means here leading-order in a
skeleton expansion, and not lowest-order in perturbation theory. In fact, the
scalar kernel $s_p$ and the vector kernel $W$ are identically zero in any order
perturbation theory.

\begin{figure}[tb]
\centering
\includegraphics[width=.5\linewidth]{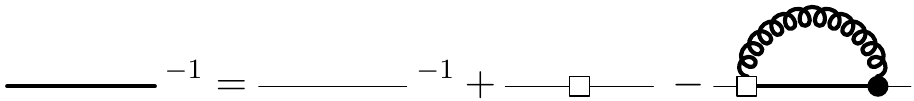}
\caption{Diagrammatic representation of the CRDSE~\eqref{qdse2} for the quark
propagator. Full lines and filled dots represent, respectively, dressed propagators
and vertices. The line with an empty square stands for the biquark kernel
$\bar\gamma$~[\Eqref{ans10a}]; the vertex with a square box represents the bare
quark-gluon vertex $\bar{\Gamma}_0$~[\Eqref{qgv0}].}
\label{fig-dse-quark}
\end{figure}

\begin{figure}[tb]
\centering
\includegraphics[scale=.7]{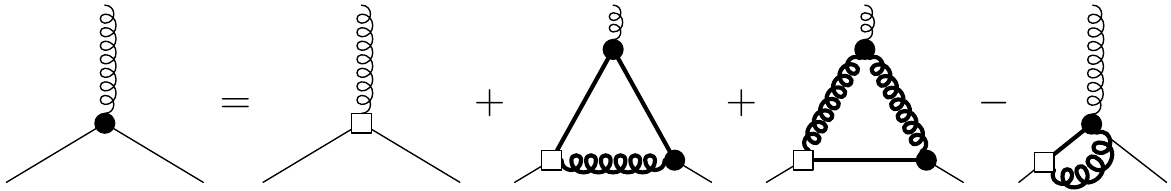}\par\bigskip
\includegraphics[scale=.7]{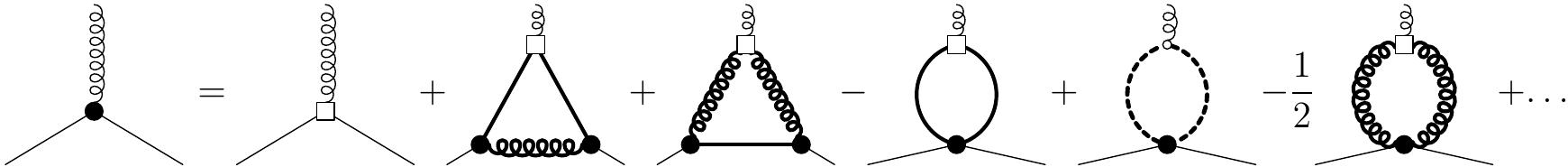}
\caption{Possible forms of the CRDSE for the quark-gluon vertex with the bare
vertex attached to the external quark (top) and gluon (bottom) leg. The ellipsis in the
bottom figure stands for two-loop diagrams.}
\label{fig-qgv-dse}
\end{figure}

In terms of the fermionic fields $\xi$, $\xi^\dag$, the physical quark propagator \Eqref{qp1}
reads
\begin{equation}\label{qp2}
S(1,2) = \vev{\xi(1) \, \xi^\dag(2)} - S_0(1,2)
\end{equation}
where $S_0$ is the bare quark propagator \Eqref{evg4}. The additional term $S_0$
is a consequence of the chosen coherent-state representation, \Eqref{csr1}, of the
fermion field operators.\footnote{%
   Other functional representations, see e.g.~Refs.~\cite{Floreanini:1987gr,Kiefer:1993fw}, would
   not give rise to such a contact term; they would, however, make the choice for the Ansatz of
   the quark vacuum wave functional much
   more involved, since one should discriminate after the minimization of the energy between the
   positive/negative energy components. The choice \Eqref{csr1} already takes care of the correct
   filling of the Dirac sea and is for our calculations much more convenient.}
The quantity
\begin{equation}\label{fprop}
Q(1,2) \equiv \vev{\xi(1)\xi^\dag(2)},
\end{equation}
referred to as fermion propagator, obeys the CRDSE \cite{Campagnari:2015zsa}
\begin{equation}\label{qdse2}
Q^{-1}(1,2) = \bigl[ 2 S_0(1,2)\bigr]^{-1} + \bar\gamma(1,2) - \bar\Gamma_0(1,3;4) Q(3,3') D(4,4') \bar\Gamma(3',2;4') ,
\end{equation}
which is diagrammatically represented in Fig.~\ref{fig-dse-quark}.
In \Eqref{qdse2}, $S_0$ is the bare quark propagator \Eqref{evg4},
\begin{equation}\label{gluonprop}
D(1,2) = \vev{A(1) A(2)}
\end{equation}
is the gluon propagator, and $\bar\Gamma$ is the full quark-gluon vertex defined by
\begin{equation}\label{qgv}
\vev{\xi(1) \xi^\dag(2) A(3)} = - Q(1,1') \, \bar\Gamma(1',2';3') \, Q(2',2) \, D(3',3) .
\end{equation}
The full quark-gluon vertex $\bar\Gamma$ also obeys a CRDSE, whose form is
diagrammatically represented in Fig.~\ref{fig-qgv-dse}.
The leading term is given by $\bar\Gamma_0$ [\Eqref{qgv0}],
thus justifying its interpretation as bare quark-gluon vertex.

Assuming a power law behaviour
\begin{equation}\label{irpow}
Q(p)\overset{p\to0}{\sim} \frac{1}{p^\delta} , \qquad \bar\Gamma(p,p)\overset{p\to0}{\sim} \frac{1}{p^\eta}
\end{equation}
for the vertex and the propagator an IR analysis of the quark
propagator CRDSE \eqref{qdse2} leads to the condition (see the Appendix)
\[
\delta=\min(0,5-\delta-\eta) ,
\]
while the vertex CRDSE truncated to include only the triangle diagrams yields
\[
\eta=\max(-1,2\eta+\delta-7,2\eta+2\delta-5) .
\]
Combining these two relations results in the following conditions for the IR exponents
\[
\begin{system}
\delta=0 \\ \eta=-1\text{ or }5
\end{system}
\qquad\text{or}\qquad
\begin{system}
-2\leq \delta \leq0 \\ \eta=5-2\delta
\end{system}
\]
In the case of an IR constant quark propagator, $\delta=0$, the possible
outcomes are $\eta=-1$ (i.e.~IR suppressed quark-gluon vertex) and $\eta=5$. In view of the experience
from FRG and DSE investigations in Landau gauge
\cite{Alkofer:2008tt,%
Aguilar:2010cn,Binosi:2016wcx,Aguilar:2014lha,Aguilar:2016lbe,Aguilar:2018epe,%
Mitter:2014wpa,Braun:2014ata,Cyrol:2017ewj,%
Windisch:2012de,Williams:2015cvx,Sanchis-Alepuz:2015qra,Oliveira:2018fkj,Oliveira:2018ukh}
such a strongly IR divergent spatial vertex
appears quite unlikely. We are confident that the solution with $\delta=0$ and
$\eta=-1$ represents the physically realized case. For this solution the full
quark-gluon vertex has the same infrared behaviour as the bare vertex.


\section{Truncation Scheme and Renormalization}
\label{sec-trunc}

\subsection{Bare-Vertex Approximation}\label{subsec:qpdse}

By global colour invariance the quark propagator has to be colour diagonal.
For the inverse quark propagator we assume the following Dirac structure
\begin{equation}\label{qdse5}
Q^{-1}(\vp) = A(\vp) \, \valpha\cdot\uvp + \beta \,  B(\vp) .
\end{equation}
In principle there might be further Dirac structures proportional to $\beta\alpha_i$
and $\id$. However, from continuum studies \cite{Watson:2011kv} it is known that
the term $\propto\beta\alpha_i$ cannot arise until two-loop order in perturbation theory.
Furthermore, lattice simulations \cite{Burgio:2012ph,Pak:2015dxa} show no indication
for the presence of such a structure.
A term proportional to the unity matrix might well exist in the full (i.e.~energy
dependent) quark propagator but drops out when performing the energy integration
required to obtain the static propagator. We will therefore stick to the form \Eqref{qdse5}.
With the explicit form of the biquark kernel \Eqref{ans10a} we obtain from the
CRDSE~\eqref{qdse2} in the chiral limit the following system of coupled equations
for the dressing functions $A$ and $B$ of the quark propagator \Eqref{qdse5}
\begin{equation}
\begin{split}
\label{1010-17a}
A_p &= 1 - 
\frac{1}{4 \Nc} \int\dfr[3]{q} \: \tr[ \valpha\cdot\uvp \bar\Gamma_{0,i}^{mn,a}(\vp,-\vq;\vq-\vp) \, Q(\vq)\, D_{ij}(\vp-\vq) \, \bar\Gamma_{j}^{nm,a}(\vq,-\vp;\vp-\vq) ] , \\
B_p &=  s_p
- \frac{1}{4\Nc} \int\dfr[3]{q} \: \tr[\beta \bar\Gamma_{0,i}^{mn,a}(\vp,-\vq;\vq-\vp) \, Q(\vq)\, D_{ij}(\vp-\vq) \, \bar\Gamma_{j}^{nm,a}(\vq,-\vp;\vp-\vq) ], \\
\end{split}
\end{equation}
where the traces are taken over Dirac indices, and
\[
D_{ij}(\vp) \equiv \frac{t_{ij}(\vp)}{2 \Omega(\vp)} , \qquad t_{ij}(\vp) = \delta_{ij} - \frac{p_i p_j}{\vp^2}
\]
is the static gluon propagator \Eqref{gluonprop}, conveniently expressed by the quasi-gluon energy $\Omega(\vp)$.

Equation \eqref{1010-17a} still involves the full quark-gluon vertex $\bar\Gamma$ [\Eqref{qgv}].
In the continuum QCD studies in Landau gauge both in the DSE and in the FRG approaches
\cite{Alkofer:2008tt,%
Aguilar:2010cn,Binosi:2016wcx,Aguilar:2014lha,Aguilar:2016lbe,Aguilar:2018epe,%
Mitter:2014wpa,Braun:2014ata,Cyrol:2017ewj,%
Windisch:2012de,Williams:2015cvx,Sanchis-Alepuz:2015qra,Oliveira:2018fkj,Oliveira:2018ukh}
a non-perturbative dressing of the quark-gluon vertex is crucial
to obtain spontaneous breaking of chiral symmetry. In Coulomb gauge the situation
is different: chiral symmetry breaking is already triggered by the instantaneous
colour Coulomb potential (see Sec.~\ref{sec:sk}) when the BCS-type wave functional
is used for the quarks, i.e.~when the vector kernels $V$ and $W$ in \Eqref{vk1}
are put to zero and thus the coupling of the quarks to the spatial gluons is
disregarded. Furthermore, it was shown in Ref.~\cite{Campagnari:2016wlt}
that for reasonable values of the strong coupling constant the inclusion of the
coupling of the quarks to the spatial gluons influences only the high-momentum
behaviour of the scalar kernel. Moreover, as shown above, bare and dressed quark-gluon
vertices have the same IR behaviour. Therefore, in the following we replace the
full quark-gluon vertex $\bar\Gamma$ by the bare one $\bar\Gamma_0$ [\Eqref{qgv0}].
We will discuss the quality of this approximation later.
After replacing the full vertices in the CRDSE \eqref{1010-17a} by bare ones,
the Dirac traces can be worked out and the coupled equations \eqref{1010-17a}
for the dressing functions of the quark propagator reduce to
\begin{subequations}\label{ndse1}
\begin{align}
\label{ndse1a}
A_p &= 1 + \frac{g^2 C_F}{2} \int \dfr[3]{q} \: \frac{A_q}{\Omega(\vp+\vq) \, \Delta_q}
\biggl[ X_-(\vp,\vq) \, V^2(\vp,\vq) + X_+(\vp,\vq) \, W^2(\vp,\vq)  \biggr] \equiv 1 + I_A , \\
\label{ndse1b}
B_p &= s_p + \frac{g^2 C_F}{2} \int \dfr[3]{q} \, \frac{B_q}{\Omega(\vp+\vq) \, \Delta_q}
\biggl[ X_-(\vp,\vq) \, V^2(\vp,\vq) - X_+(\vp,\vq) \, W^2(\vp,\vq)  \biggr] ,
\end{align}
\end{subequations}
where $C_F=(\Nc^2-1)/(2\Nc)$ is the Casimir eigenvalue in the fundamental representation,
\[
\Delta_p = A_p^2 + B_p^2,
\]
and the factors
\[
X_\pm(\vp,\vq) = 1 \pm \frac{[\uvp\cdot(\vp+\vq)] [\uvq\cdot(\vp+\vq)]}{(\vp+\vq)^2}
\]
arise from the contraction of the trace of Dirac matrices with the transverse
projector $t_{ij}(\vp)$.

\subsection{The Biquark Kernel}
\label{sec:sk}

\begin{figure}
\includegraphics[width=.4\linewidth]{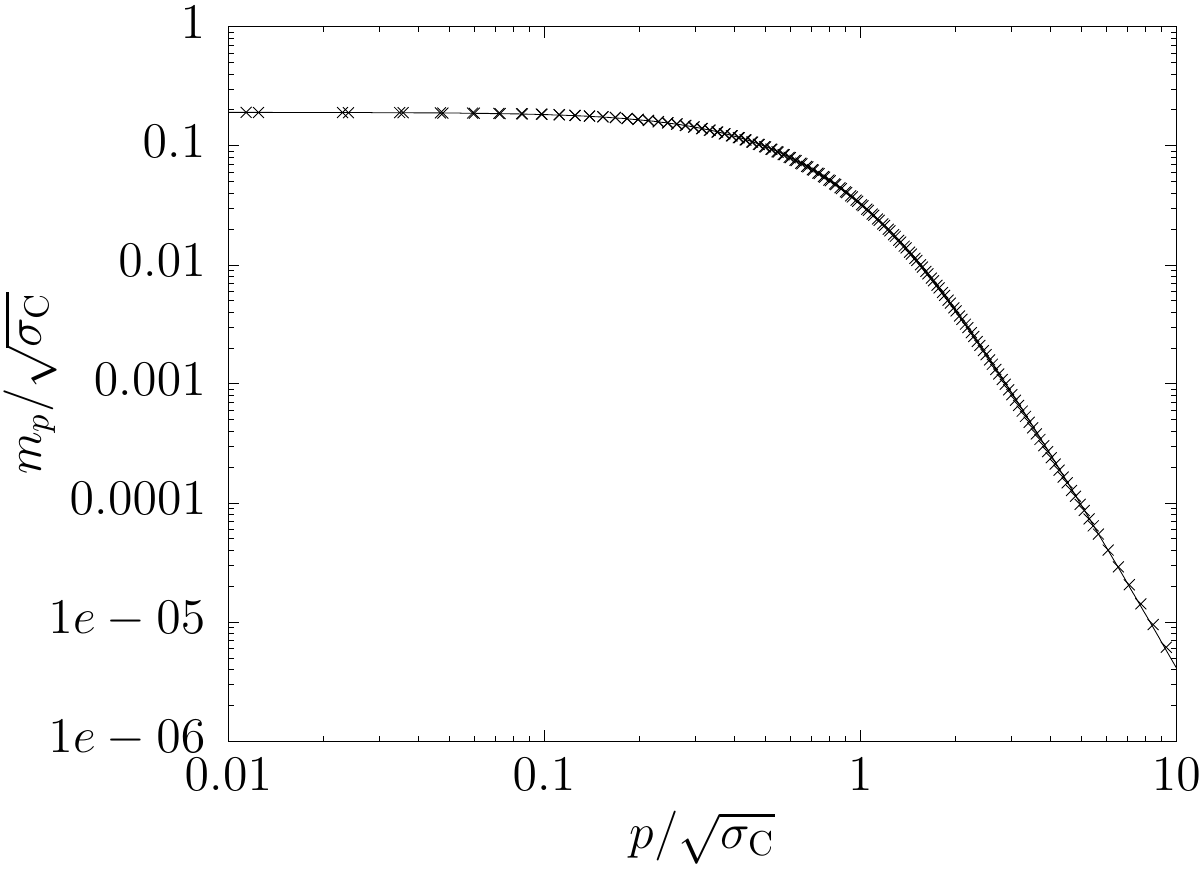}
\caption{Numerical solution of the gap equation \eqref{ad2} together with the fit \Eqref{ndse4}.}
\label{fig-ad}
\end{figure}

In principle we should solve the two CRDSEs \eqref{ndse1} with the vector
kernels $V$ and $W$ [\Eqref{vk}] together with the gap equation for the scalar
kernel $s_p$ derived in Refs.~\cite{Vastag:2015qjd,Campagnari:2016wlt,Campagnari:2018flz}. However,
these previous investigations have shown that the effect of the spatial gluons on
the scalar kernel~$s_p$ is rather small. For physical values of the coupling~$g$,
the scalar kernel is dominated by the strongly IR divergent colour Coulomb potential:
in particular the IR behaviour of $s_p$ is exclusively determined by the Coulomb
term. Therefore, in the following we assume that the scalar kernel~$s_p$ satisfies
the gap equation obtained without the coupling to the spatial gluons, which
is given by
\begin{equation}\label{ad}
\abs{\vp} \, s_p = \frac{g^2 C_F}{2} \int \dfr[3]{q} \, F(\vp-\vq) \, \frac{s_q(1-s_p^2) - \uvp\cdot\uvq \, s_p(1-s^2_q)}{1+s_q^2} .
\end{equation}
Here, the colour Coulomb potential $F(\vp)$ is the expectation value of the
Coulomb kernel \Eqref{coulkernel}.\footnote{%
   When $F(\vp)$ is replaced by a linearly rising potential or by more general
   types of four-body instantaneous potentials one recovers the equations studied
   in Refs~\cite{Finger:1979yt,Finger:1981gm,Amer:1983qa,LeYaouanc:1983huv,Adler:1984ri,Davis:1984xg,Alkofer:1988tc,Bicudo:1989sh}.}
Previous studies
\cite{Zwanziger:2002sh,Zwanziger:2003de,Epple:2006hv,Burgio:2015hsa}
show that in coordinate space the expectation value of $F_A$ raises linearly for
large distances. In fact, an IR analysis of the variational equations of the
Yang--Mills sector reveals the IR behaviour
\begin{equation}\label{coulprop}
g^2 F(\vp) = \frac{8\pi\sigmacoul}{(\vp^2)^2},
\end{equation}
where $\sigmacoul$ is the Coulomb string tension: $\sigmacoul$
is larger than the Wilson string tension $\sigma$ by a factor ranging from 2.5 to 4
\cite{Nakagawa:2006fk,Golterman:2012dx,Greensite:2014bua}, the
latter value seemingly being favoured by recent lattice calculations \cite{Greensite:2015nea}.
In the numerical solution of \Eqref{ad} we use the IR from \Eqref{coulprop} of
the Coulomb propagator
and set $\sigmacoul\simeq4\sigma$, fixing our scale at $\sqrt{\sigmacoul}=0.88\,\mathrm{GeV}$.
For numerical reasons the gap equation \eqref{ad} is reformulated in terms of
the mass function\footnote{%
   Let us stress that the interpretation of $m_p$ as `mass function' of the quark
   propagator is valid only when one ignores the coupling of the quark field to
   the spatial gluons and sets $A_p=1$ and $B_p=s_p$. When the interaction
   with the spatial gluons is taken into account, $m_p$ is just a useful
   auxiliary quantity.}
$m_p$
\begin{equation}\label{ndse3}
m_p = \frac{2ps_p}{1-s_p^2} \implies s_p = \sqrt{1+p^2/m_p^2} - p/m_p ,
\end{equation}
resulting in
\begin{equation}\label{ad2}
m_p = \frac{g^2 C_F}{2} \int \dfr[3]{q} \frac{F(\vp-\vq)}{\sqrt{\vq^2+m_q^2}} \biggl[m_q - \frac{\vp\cdot\vq}{\vp^2} \,  m_p \biggr] .
\end{equation}
The numerical solution \cite{Quandt:2018bbu} of the gap equation \Eqref{ad2} can be fitted by
\begin{equation}\label{ndse4}
m_p = \frac{0.19099 \sqrt{\sigmacoul}}{\bigl[ 1 + 0.95086 \bigl(p/\sqrt{\sigmacoul}\bigr){}^{1.7648} \bigr]^{2.6632}} .
\end{equation}
Figure~\ref{fig-ad} shows the numerical solution of the gap equation \eqref{ad2}
together with the fit \Eqref{ndse4}.

\subsection{Renormalization and Chiral Condensate}

The equation \eqref{ndse1a} for the dressing function $A_p$ is UV divergent. Using
a sharp UV cut-off we find
\[
A_p = 1 + \frac{\alpha_s C_F}{4\pi} (1+s_p^2) \ln\Lambda + \text{finite terms}
\]
where $\alpha_s=g^2/(4\pi)$ is the strong coupling constant. As we have shown in
Ref.~\cite{Campagnari:2018flz}, the occurrence of a momentum-dependent divergence
poses no conceptual problem, since $A_p$ and $B_p$ are the dressing functions of
the propagator $Q$ [\Eqref{fprop}] of the coherent-state fields $\xi$, $\xi^\dag$
and not of the physical quark propagator $S$ [Eqs.~\eqref{qp1} and \eqref{qp2}].
We renormalize the equation for $A_p$ by subtracting
it at an arbitrary scale $\mu$, obtaining
\begin{equation}\label{ndse2}
A_p = 1 + \frac{1+s_p^2}{1+s_\mu^2} (A_\mu-1) + \lim_{\Lambda\to\infty} \biggl[ I_A(p,\Lambda) - \frac{1+s_p^2}{1+s_\mu^2} \, I_A(\mu,\Lambda) \biggr] ,
\end{equation}
where $I_A$ is the loop integral on the right-hand side of \Eqref{ndse1a}.

Once the results for $A_p$ and $B_p$ are known, the chiral condensate can be evaluated by
\[
\vev{\bar{q}q} = - \frac{2\Nc}{\pi^2} \int \d p \, p^2 \frac{B_p}{A_p^2+B_p^2} .
\]


\section{Results}
\label{sec-res}
\begin{figure*}
\parbox{.45\linewidth}{\centering\includegraphics[width=\linewidth]{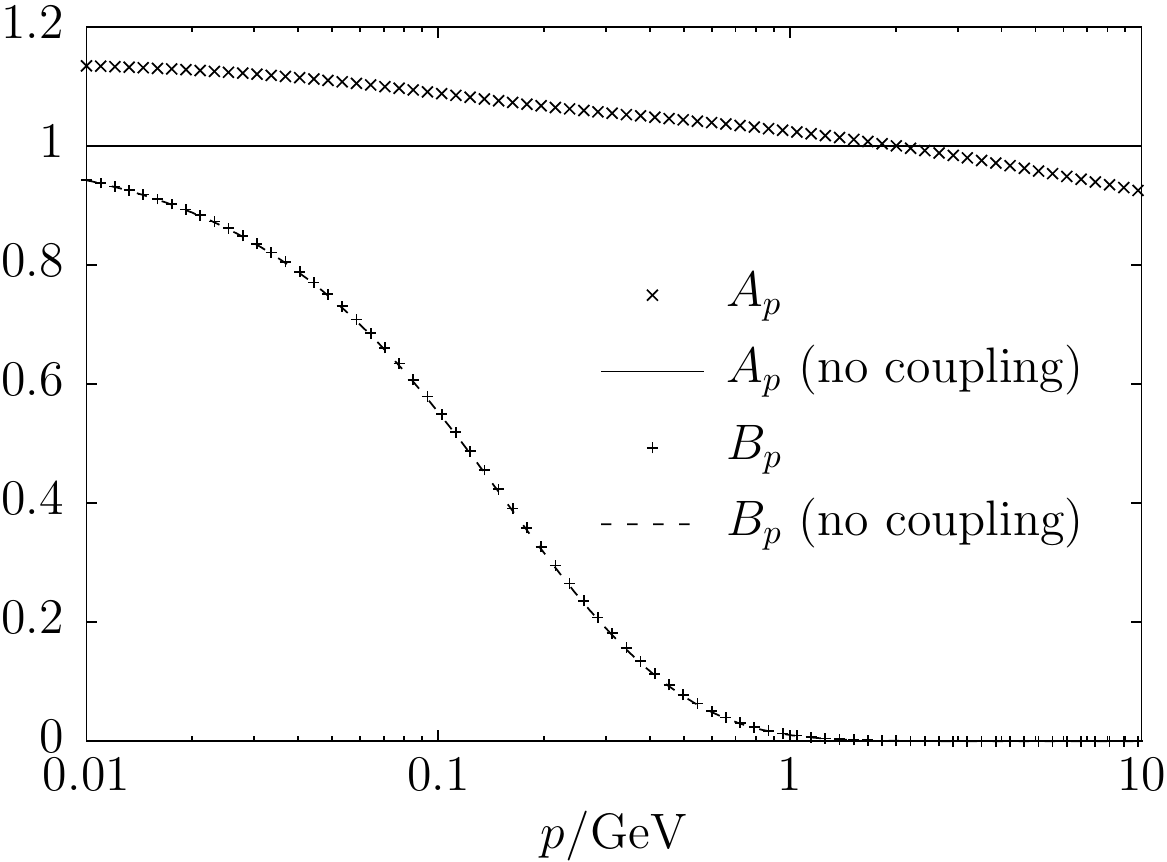}\\(a)}\hfill
\parbox{.45\linewidth}{\centering\includegraphics[width=\linewidth]{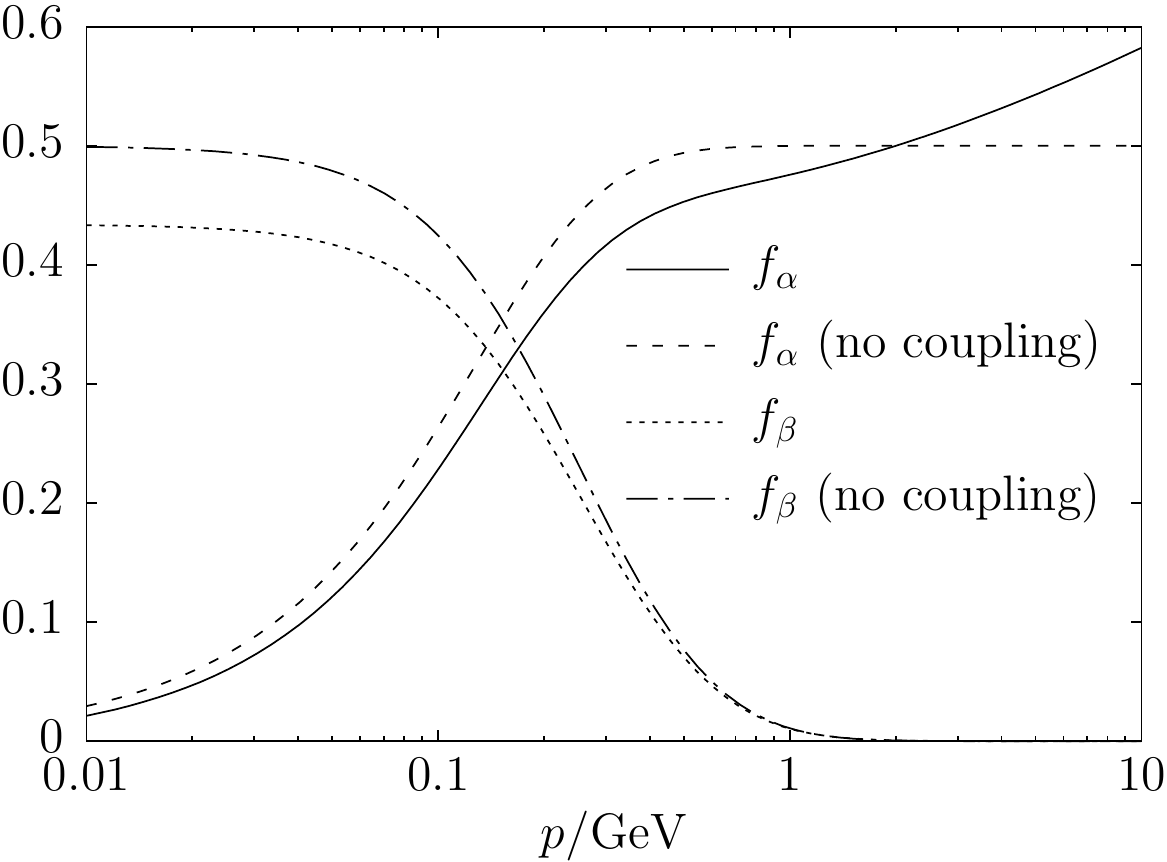}\\(b)}
\caption{Dressing functions $A_p$ and $B_p$ of the fermion propagator \Eqref{fprop}
renormalized at $\mu=2\,\mathrm{GeV}$ (left panel) and of the physical quark
propagator \Eqref{qp3} (right panel). The curves labelled `no coupling' refer to
the case where the coupling to the spatial gluons is ignored, iwhich results in
$A_p=1$ and $B_p=s_p$.}
\label{fig:res1}
\end{figure*}

We have solved the coupled equations \eqref{ndse1b} and \eqref{ndse2} with the variational kernels
$V$ [\Eqref{vkV}], $W$ [\Eqref{vkW}] and $s_p$ [\Eqref{ndse3}] calculated
from the solution \Eqref{ndse4} of the gap equation \eqref{ad}.
The quasi-gluon energy $\Omega$ is parametrized by the Gribov formula \cite{Gribov:1977wm,Burgio:2008jr}
\begin{equation}\label{gribov}
\Omega(\vp) = \sqrt{\vp^2 + m_A^4/\vp^2}.
\end{equation}
For the Gribov mass $m_A$ we have taken the value
\[
m_A^2 = \frac{\Nc}{\pi} \, \sigmacoul
\]
resulting from the IR analysis of the gluon gap equation \cite{Epple:2006hv}.
The results are shown in Fig.~\ref{fig:res1} in a MOM scheme with $A_\mu=1$ for
the renormalization scale $\mu=2\,\mathrm{GeV}$.
The value $\alpha_s(2\,\mathrm{GeV})=0.30(1)$ \cite{Buckley:2014ana} in the
$\overline{\text{MS}}$ scheme can be converted to the MOM value
$\alpha_s^{\mathrm{MOM}}(2\,\mathrm{GeV})\simeq0.44$ by means of the three-loop
$\beta$ function \cite{Chetyrkin:2000fd}.\footnote{%
   In Ref.~\cite{Aguilar:2014lha} the coupling was fixed to $\alpha_s^{\mathrm{MOM}}(2\,\mathrm{GeV})=0.45$
   by fitting the ghost propagator to the lattice data.}
Figure~\ref{fig:res1}a shows the dressing functions $A_p$ and $B_p$ of the fermion
propagator \Eqref{fprop} together with the values $A_p=1$ and $B_p=s_p$ which
follow when the coupling to the spatial gluons is ignored. The physical quark
propagator \Eqref{qp2} becomes in momentum space
\begin{equation}\label{qp3}
S(\vp) = Q(\vp) - S_0(\vp) = \frac{A_p \valpha\cdot\uvp + B_p \beta}{A_p^2+B_p^2} - \frac{\valpha\cdot\uvp}{2}
\equiv f_\alpha(p) \valpha\cdot\uvp + f_\beta(p) \beta .
\end{equation}
with dressing functions
\[
f_\alpha(p) = \frac{A_p}{A_p^2+B_p^2} - \frac{1}{2} , \qquad
f_\beta(p) = \frac{B_p}{A_p^2+B_p^2} ,
\]
which are shown in Fig.~\ref{fig:res1}b.
The dressing function $f_\alpha$ of the $\alpha_i$ piece of the propagator
develops an anomalous dimension in the UV, compatible with the perturbative
analysis \cite{Campagnari:2014hda}.

For the chiral condensate we recover the value $(-236\,\mathrm{MeV})^3$. Both
the dressing functions of the physical quark propagator and the chiral condensate
differ only little from the values obtained by ignoring the coupling to the
spatial gluons and from the one-loop expansion of Ref.~\cite{Campagnari:2018flz}.


\section{Conclusions}
\label{sec-conc}

We have solved numerically the CRDSEs for the quark propagator under the assumptions
that i\rparen\ the scalar kernel is dominated by the IR diverging colour Coulomb
interactions, and ii\rparen\ that we can replace the full quark-gluon vertex by
the bare one. While we are very confident that the first assumption is reliable,
the second one is less under control. It is well-known from Landau
gauge studies
\cite{Alkofer:2008tt,%
Aguilar:2010cn,Binosi:2016wcx,Aguilar:2014lha,Aguilar:2016lbe,Aguilar:2018epe,%
Mitter:2014wpa,Braun:2014ata,Cyrol:2017ewj,%
Windisch:2012de,Williams:2015cvx,Sanchis-Alepuz:2015qra,Oliveira:2018fkj,Oliveira:2018ukh}
that a number of Dirac structures beyond the leading order
$\gamma^\mu$ contribute to the overall strength of the vertex. The
quark-gluon vertex in Coulomb gauge is currently under investigation. It is more
involved than its Landau-gauge counterpart: because of the projectors \Eqref{qptproj3} in \Eqref{qgv0},
the bare vertex alone involves five different Dirac structures instead of one. Furthermore,
since the Coulomb gauge is non-covariant the total number of Dirac structures
must be nearly doubled, because the temporal and spatial Dirac matrices must be
treated separately. It is very likely
that all these further tensor structures appearing in the full vertex are not
as relevant as in Landau gauge; this is because in our approach chiral symmetry
breaking is caused by the Coulomb term and not by the quark-gluon vertex as it
is the case in Landau gauge.
Keeping only this dominant contribution
could be a useful phenomenological tool for further calculations at finite
temperature and density.


\begin{acknowledgments}
This work was supported but the Deutsche Forschungsgemeinschaft (DFG) under contract
No.~DFG-Re856/10-1.
\end{acknowledgments}


\appendix

\section{Infrared Analysis}
Suppressing all indices, the fermion propagator CRDSE \eqref{qdse2} can be written
as
\begin{equation}\label{ira1}
Q^{-1} = \bigl[ 2S_0 \bigr]^{-1} + \bar\gamma + \int \bar\Gamma Q \bar\Gamma_0 D .
\end{equation}
The IR analysis is performed in the usual way by rescaling all momenta by a factor
$\lambda$, $\vp\to\lambda\vp$ and $\vq\to\lambda\vq$, sending $\lambda\to0$,
replacing the various Green functions by their assumed IR scaling behaviour [\Eqref{irpow}], and
comparing the powers of $\lambda$. The biquark kernel $\bar\gamma$ [Eqs.~\eqref{ans10a} and \eqref{sk1}]
and the bare quark propagator $S_0$ [\Eqref{evg4}] are IR constant
\[
\bar\gamma \sim S_0 \sim \lambda^0
\]
while the bare quark-gluon vertex [Eqs.~\eqref{qgv0} and \eqref{vk1}] vanishes in
the IR because of the IR diverging gluon energy $\Omega$ [\Eqref{gribov}] in
the denominator of the vector kernels $V$ and $W$ [Eqs.~\eqref{vk}]
\[
\bar\Gamma_0 \sim \lambda .
\]
From \Eqref{ira1} we find with a factor $\lambda^3$ coming from the momentum integration:
\[
\lambda^\delta \sim \lambda^0 + \lambda^{3-\eta-\delta+1+1} .
\]
For $\lambda\to0$ the smaller exponent dominates, and we obtain
\[
\delta = \min(0,5-\delta-\eta) .
\]
For the IR analysis of the CRDSE for the quark-gluon vertex we choose the equation
with the bare vertex attached to the external gluon leg (second equation in Fig.~\ref{fig-qgv-dse})
and keep only the triangle diagrams. This CRDSE then becomes
\begin{equation}\label{fqgve}
\bar\Gamma = \bar\Gamma_0 + \int \bar\Gamma Q \bar\Gamma_0 Q \bar\Gamma D + \int \bar\Gamma Q \bar\Gamma D^2 \gamma_3
\end{equation}
where $\gamma_3$ is the three-gluon kernel determined in Ref.~\cite{Campagnari:2010wc}.
The latter vanishes as $\lambda^2$ for $\lambda\to0$.
From the CRDSE~\eqref{fqgve} we find
\[
\lambda^{-\eta} \sim \lambda + \lambda^{3-\eta-\delta+1-\delta-\eta+1} + \lambda^{3-\eta-\delta-\eta+2+2}
\]
resulting in
\[
-\eta = \min(1,5-2\delta-2\eta,7-2\eta-\delta) .
\]
Using the version of the CRDSE with the bare vertex attached to the incoming quark line
(first equation in Fig.~\ref{fig-qgv-dse})
forces one to consider also the full three-gluon vertex, which was investigated
in Ref.~\cite{Huber:2014isa}. However, the conclusions presented at the end of
Sec.~\ref{sec-def} remain unaltered.

A possibly strong IR enhancement might come from the quark four-point function
(see Fig.~\ref{fig-qgv-dse} bottom). Even if our vacuum wave functional \Eqref{vwf}
does not involve a four-point function, we can nevertheless try to estimate its
effect: if we had such a term in our Ansatz, its leading-order expansion would
be of the form
\[
\bar{\Gamma}_{0,4q} \sim F + \bar{\Gamma}_{0,i} \bar{\Gamma}_{0,i}
\]
where $F$ is the Coulomb propagator \Eqref{coulprop}. Even with an IR exponent
of $-4$ from the Coulomb propagator the whole diagram would not change the
results of the IR analysis.


\bibliographystyle{h-physrev5}
\bibliography{biblio-spires}

\begin{thebibliography}{10}

\bibitem{Alkofer:2000wg}
R.~Alkofer and L.~von Smekal,
\newblock Phys. Rept. {\bfseries 353}, 281 (2001), arXiv:hep-ph/0007355.

\bibitem{Pawlowski:2005xe}
J.~M. Pawlowski,
\newblock Annals Phys. {\bfseries 322}, 2831 (2007), arXiv:hep-th/0512261.

\bibitem{Fischer:2006ub}
C.~S. Fischer,
\newblock J. Phys. {\bfseries G32}, R253 (2006), arXiv:hep-ph/0605173.

\bibitem{Binosi:2009qm}
D.~Binosi and J.~Papavassiliou,
\newblock Phys. Rept. {\bfseries 479}, 1 (2009), arXiv:0909.2536.

\bibitem{Braun:2011pp}
J.~Braun,
\newblock J. Phys. {\bfseries G39}, 033001 (2012), arXiv:1108.4449.

\bibitem{Eichmann:2016yit}
G.~Eichmann, H.~Sanchis-Alepuz, R.~Williams, R.~Alkofer, and C.~S. Fischer,
\newblock Prog. Part. Nucl. Phys. {\bfseries 91}, 1 (2016), arXiv:1606.09602.

\bibitem{Aoki:2016frl}
S.~Aoki {\em et~al.},
\newblock Eur. Phys. J. {\bfseries C77}, 112 (2017), arXiv:1607.00299.

\bibitem{Nambu:1974zg}
Y.~Nambu,
\newblock Phys. Rev. {\bfseries D10}, 4262 (1974).

\bibitem{tHooft:1981bkw}
G.~'t~Hooft,
\newblock Nucl. Phys. {\bfseries B190}, 455 (1981).

\bibitem{Mack:1978rq}
G.~Mack and V.~B. Petkova,
\newblock Annals Phys. {\bfseries 123}, 442 (1979).

\bibitem{Nielsen:1979xu}
H.~B. Nielsen and P.~Olesen,
\newblock Nucl. Phys. {\bfseries B160}, 380 (1979).

\bibitem{DelDebbio:1998luz}
L.~Del~Debbio, M.~Faber, J.~Giedt, J.~Greensite, and S.~Olejnik,
\newblock Phys. Rev. {\bfseries D58}, 094501 (1998), arXiv:hep-lat/9801027.

\bibitem{Langfeld:1997jx}
K.~Langfeld, H.~Reinhardt, and O.~Tennert,
\newblock Phys. Lett. {\bfseries B419}, 317 (1998), arXiv:hep-lat/9710068.

\bibitem{Engelhardt:1999fd}
M.~Engelhardt, K.~Langfeld, H.~Reinhardt, and O.~Tennert,
\newblock Phys. Rev. {\bfseries D61}, 054504 (2000), arXiv:hep-lat/9904004.

\bibitem{Gribov:1977wm}
V.~Gribov,
\newblock Nucl. Phys. {\bfseries B139}, 1 (1978).

\bibitem{Kugo:1979gm}
T.~Kugo and I.~Ojima,
\newblock Prog. Theor. Phys. Suppl. {\bfseries 66}, 1 (1979).

\bibitem{Zwanziger:1998ez}
D.~Zwanziger,
\newblock Nucl. Phys. {\bfseries B518}, 237 (1998).

\bibitem{Feuchter:2004mk}
C.~Feuchter and H.~Reinhardt,
\newblock Phys. Rev. {\bfseries D70}, 105021 (2004), arXiv:hep-th/0408236.

\bibitem{Epple:2006hv}
D.~Epple, H.~Reinhardt, and W.~Schleifenbaum,
\newblock Phys. Rev. {\bfseries D75}, 045011 (2007), arXiv:hep-th/0612241.

\bibitem{Greensite:2011zz}
J.~Greensite,
\newblock Lect. Notes Phys. {\bfseries 821}, 1 (2011).

\bibitem{Greensite:2004ur}
J.~Greensite, S.~Olejnik, and D.~Zwanziger,
\newblock JHEP {\bfseries 05}, 070 (2005), arXiv:hep-lat/0407032.

\bibitem{Reinhardt:2008ek}
H.~Reinhardt,
\newblock Phys. Rev. Lett. {\bfseries 101}, 061602 (2008), arXiv:0803.0504.

\bibitem{Burgio:2015hsa}
G.~Burgio, M.~Quandt, H.~Reinhardt, and H.~Vogt,
\newblock Phys. Rev. {\bfseries D92}, 034518 (2015), arXiv:1503.09064.

\bibitem{Schutte:1985sd}
D.~Schutte,
\newblock Phys. Rev. {\bfseries D31}, 810 (1985).

\bibitem{Szczepaniak:2001rg}
A.~P. Szczepaniak and E.~S. Swanson,
\newblock Phys. Rev. {\bfseries D65}, 025012 (2001), arXiv:hep-ph/0107078.

\bibitem{Campagnari:2010wc}
D.~R. Campagnari and H.~Reinhardt,
\newblock Phys. Rev. {\bfseries D82}, 105021 (2010), arXiv:1009.4599.

\bibitem{Pak:2011wu}
M.~Pak and H.~Reinhardt,
\newblock Phys. Lett. {\bfseries B707}, 566 (2012), arXiv:1107.5263.

\bibitem{Pak:2013uba}
M.~Pak and H.~Reinhardt,
\newblock Phys. Rev. {\bfseries D88}, 125021 (2013), arXiv:1310.1797.

\bibitem{Campagnari:2015zsa}
D.~R. Campagnari and H.~Reinhardt,
\newblock Phys. Rev. {\bfseries D92}, 065021 (2015), arXiv:1507.01414.

\bibitem{Vastag:2015qjd}
P.~Vastag, H.~Reinhardt, and D.~Campagnari,
\newblock Phys.~Rev. {\bfseries D93}, 065003 (2016), arXiv:1512.06733.

\bibitem{Reinhardt:2016pfe}
H.~Reinhardt and P.~Vastag,
\newblock Phys. Rev. {\bfseries D94}, 105005 (2016), arXiv:1605.03740.

\bibitem{Campagnari:2016wlt}
D.~R. Campagnari, E.~Ebadati, H.~Reinhardt, and P.~Vastag,
\newblock Phys.~Rev. {\bfseries D94}, 074027 (2016), arXiv:1608.06820.

\bibitem{Campagnari:2018flz}
D.~Campagnari and H.~Reinhardt,
\newblock Phys. Rev. {\bfseries D97}, 054027 (2018), arXiv:1801.02045.

\bibitem{Quandt:2013wna}
M.~Quandt, H.~Reinhardt, and J.~Heffner,
\newblock Phys. Rev. {\bfseries D89}, 065037 (2014), arXiv:1310.5950.

\bibitem{Quandt:2015aaa}
M.~Quandt and H.~Reinhardt,
\newblock Phys. Rev. {\bfseries D92}, 025051 (2015), arXiv:1503.06993.

\bibitem{Christ:1980ku}
N.~H. Christ and T.~D. Lee,
\newblock Phys. Rev. {\bfseries D22}, 939 (1980).

\bibitem{Mansfield:1997kj}
P.~Mansfield, M.~Sampaio, and J.~Pachos,
\newblock Int. J. Mod. Phys. {\bfseries A13}, 4101 (1998),
  arXiv:hep-th/9702072.

\bibitem{Mansfield:1998vc}
P.~Mansfield and M.~Sampaio,
\newblock Nucl. Phys. {\bfseries B545}, 623 (1999), arXiv:hep-th/9807163.

\bibitem{Campagnari:2009km}
D.~R. Campagnari, H.~Reinhardt, and A.~Weber,
\newblock Phys. Rev. {\bfseries D80}, 025005 (2009), arXiv:0904.3490.

\bibitem{Campagnari:2009wj}
D.~Campagnari, A.~Weber, H.~Reinhardt, F.~Astorga, and W.~Schleifenbaum,
\newblock Nucl. Phys. {\bfseries B842}, 501 (2011), arXiv:0910.4548.

\bibitem{Campagnari:2014hda}
D.~R. Campagnari and H.~Reinhardt,
\newblock Int.~J.~Mod.~Phys. {\bfseries A30}, 1550100 (2015), arXiv:1404.2797.

\bibitem{Floreanini:1987gr}
R.~Floreanini and R.~Jackiw,
\newblock Phys. Rev. {\bfseries D37}, 2206 (1988).

\bibitem{Kiefer:1993fw}
C.~Kiefer and A.~Wipf,
\newblock Annals Phys. {\bfseries 236}, 241 (1994), arXiv:hep-th/9306161.

\bibitem{Alkofer:2008tt}
R.~Alkofer, C.~S. Fischer, F.~J. Llanes-Estrada, and K.~Schwenzer,
\newblock Annals Phys. {\bfseries 324}, 106 (2009), arXiv:0804.3042.

\bibitem{Aguilar:2010cn}
A.~Aguilar and J.~Papavassiliou,
\newblock Phys. Rev. {\bfseries D83}, 014013 (2011), arXiv:1010.5815.

\bibitem{Binosi:2016wcx}
D.~Binosi, L.~Chang, J.~Papavassiliou, S.-X. Qin, and C.~D. Roberts,
\newblock Phys. Rev. {\bfseries D95}, 031501 (2017), arXiv:1609.02568.

\bibitem{Aguilar:2014lha}
A.~C. Aguilar, D.~Binosi, D.~Ibañez, and J.~Papavassiliou,
\newblock Phys. Rev. {\bfseries D90}, 065027 (2014), arXiv:1405.3506.

\bibitem{Aguilar:2016lbe}
A.~C. Aguilar, J.~C. Cardona, M.~N. Ferreira, and J.~Papavassiliou,
\newblock Phys. Rev. {\bfseries D96}, 014029 (2017), arXiv:1610.06158.

\bibitem{Aguilar:2018epe}
A.~C. Aguilar, J.~C. Cardona, M.~N. Ferreira, and J.~Papavassiliou,
\newblock Phys. Rev. {\bfseries D98}, 014002 (2018), arXiv:1804.04229.

\bibitem{Mitter:2014wpa}
M.~Mitter, J.~M. Pawlowski, and N.~Strodthoff,
\newblock Phys. Rev. {\bfseries D91}, 054035 (2015), arXiv:1411.7978.

\bibitem{Braun:2014ata}
J.~Braun, L.~Fister, J.~M. Pawlowski, and F.~Rennecke,
\newblock Phys. Rev. {\bfseries D94}, 034016 (2016), arXiv:1412.1045.

\bibitem{Cyrol:2017ewj}
A.~K. Cyrol, M.~Mitter, J.~M. Pawlowski, and N.~Strodthoff,
\newblock Phys. Rev. {\bfseries D97}, 054006 (2018), arXiv:1706.06326.

\bibitem{Windisch:2012de}
A.~Windisch, M.~Hopfer, and R.~Alkofer,
\newblock (2012), arXiv:1210.8428,
\newblock [Acta Phys. Polon. Supp.6,347(2013)].

\bibitem{Williams:2015cvx}
R.~Williams, C.~S. Fischer, and W.~Heupel,
\newblock Phys. Rev. {\bfseries D93}, 034026 (2016), arXiv:1512.00455.

\bibitem{Sanchis-Alepuz:2015qra}
H.~Sanchis-Alepuz and R.~Williams,
\newblock Phys. Lett. {\bfseries B749}, 592 (2015), arXiv:1504.07776.

\bibitem{Oliveira:2018fkj}
O.~Oliveira, T.~Frederico, W.~de~Paula, and J.~P. B.~C. de~Melo,
\newblock Eur. Phys. J. {\bfseries C78}, 553 (2018), arXiv:1807.00675.

\bibitem{Oliveira:2018ukh}
O.~Oliveira, W.~de~Paula, T.~Frederico, and J.~P. B.~C. de~Melo,
\newblock Eur. Phys. J. {\bfseries C79}, 116 (2019), arXiv:1807.10348.

\bibitem{Watson:2011kv}
P.~Watson and H.~Reinhardt,
\newblock Phys. Rev. {\bfseries D85}, 025014 (2012), arXiv:1111.6078,
\newblock 27 pages, 11 figures.

\bibitem{Burgio:2012ph}
G.~Burgio, M.~Schrock, H.~Reinhardt, and M.~Quandt,
\newblock Phys. Rev. {\bfseries D86}, 014506 (2012), arXiv:1204.0716.

\bibitem{Pak:2015dxa}
M.~Pak and M.~Schr\"ock,
\newblock Phys. Rev. {\bfseries D91}, 074515 (2015), arXiv:1502.07706.

\bibitem{Finger:1979yt}
J.~Finger, D.~Horn, and J.~Mandula,
\newblock Phys. Rev. {\bfseries D20}, 3253 (1979).

\bibitem{Finger:1981gm}
J.~R. Finger and J.~E. Mandula,
\newblock Nucl. Phys. {\bfseries B199}, 168 (1982).

\bibitem{Amer:1983qa}
A.~Amer, A.~Le~Yaouanc, L.~Oliver, O.~Pene, and J.~c. Raynal,
\newblock Phys. Rev. Lett. {\bfseries 50}, 87 (1983).

\bibitem{LeYaouanc:1983huv}
A.~Le~Yaouanc, L.~Oliver, O.~Pene, and J.~C. Raynal,
\newblock Phys. Rev. {\bfseries D29}, 1233 (1984).

\bibitem{Adler:1984ri}
S.~L. Adler and A.~Davis,
\newblock Nucl. Phys. {\bfseries B244}, 469 (1984).

\bibitem{Davis:1984xg}
A.~C. Davis and A.~M. Matheson,
\newblock Nucl. Phys. {\bfseries B246}, 203 (1984).

\bibitem{Alkofer:1988tc}
R.~Alkofer and P.~Amundsen,
\newblock Nucl. Phys. {\bfseries B306}, 305 (1988).

\bibitem{Bicudo:1989sh}
P.~J. d.~A. Bicudo and J.~E. F.~T. Ribeiro,
\newblock Phys. Rev. {\bfseries D42}, 1611 (1990).

\bibitem{Zwanziger:2002sh}
D.~Zwanziger,
\newblock Phys. Rev. Lett. {\bfseries 90}, 102001 (2003),
  arXiv:hep-lat/0209105.

\bibitem{Zwanziger:2003de}
D.~Zwanziger,
\newblock Phys. Rev. {\bfseries D70}, 094034 (2004), arXiv:hep-ph/0312254.

\bibitem{Nakagawa:2006fk}
Y.~Nakagawa, A.~Nakamura, T.~Saito, H.~Toki, and D.~Zwanziger,
\newblock Phys. Rev. {\bfseries D73}, 094504 (2006), arXiv:hep-lat/0603010.

\bibitem{Golterman:2012dx}
M.~Golterman, J.~Greensite, S.~Peris, and A.~P. Szczepaniak,
\newblock Phys. Rev. {\bfseries D85}, 085016 (2012), arXiv:1201.4590.

\bibitem{Greensite:2014bua}
J.~Greensite and A.~P. Szczepaniak,
\newblock Phys. Rev. {\bfseries D91}, 034503 (2015), arXiv:1410.3525.

\bibitem{Greensite:2015nea}
J.~Greensite and A.~P. Szczepaniak,
\newblock Phys. Rev. {\bfseries D93}, 074506 (2016), arXiv:1505.05104.

\bibitem{Quandt:2018bbu}
M.~Quandt, E.~Ebadati, H.~Reinhardt, and P.~Vastag,
\newblock Phys. Rev. {\bfseries D98}, 034012 (2018), arXiv:1806.04493.

\bibitem{Burgio:2008jr}
G.~Burgio, M.~Quandt, and H.~Reinhardt,
\newblock Phys. Rev. Lett. {\bfseries 102}, 032002 (2009), arXiv:0807.3291.

\bibitem{Buckley:2014ana}
A.~Buckley {\em et~al.},
\newblock Eur. Phys. J. {\bfseries C75}, 132 (2015), arXiv:1412.7420.

\bibitem{Chetyrkin:2000fd}
K.~G. Chetyrkin and T.~Seidensticker,
\newblock Phys. Lett. {\bfseries B495}, 74 (2000), arXiv:hep-ph/0008094.

\bibitem{Huber:2014isa}
M.~Q. Huber, D.~R. Campagnari, and H.~Reinhardt,
\newblock Phys. Rev. {\bfseries D91}, 025014 (2015), arXiv:1410.4766.

\end{thebibliography}

\end{document}